\documentclass{article}

\usepackage{arxiv}

\usepackage[utf8]{inputenc} 
\usepackage[T1]{fontenc}    
\usepackage[hidelinks]{hyperref}  
\usepackage{url}            
\usepackage{booktabs}       
\usepackage{amsfonts}       
\usepackage{nicefrac}       
\usepackage{microtype}      	
\usepackage{graphicx}
\usepackage[numbers,square]{natbib}
\usepackage{doi}
\usepackage{graphicx}
\usepackage{pdfpages}
\usepackage{mathtools}
\usepackage{caption}
\usepackage{subcaption}
\usepackage{enumitem}
\usepackage{listings}
\usepackage{algorithm}
\usepackage[noend]{algpseudocode}
\usepackage{multirow}
\usepackage{array}
\usepackage{tabularx}
\usepackage{makecell}
\usepackage{color}
\usepackage{xcolor}
\usepackage{layouts}
\usepackage{amsthm, amssymb}
\usepackage{wrapfig}

\usepackage{url}

\DeclarePairedDelimiter\ceil{\lceil}{\rceil}

\DeclareMathOperator*{\lcm}{lcm}
\makeatletter
\renewcommand{\Function}[2]{%
  \csname ALG@cmd@\ALG@L @Function\endcsname{#1}{#2}%
  \def\jayden@currentfunction{#1}%
}
\newcommand{\funclabel}[1]{%
  \@bsphack
  \protected@write\@auxout{}{%
    \string\newlabel{#1}{{\jayden@currentfunction}{\thepage}}%
  }%
  \@esphack
}
\makeatother

\newcommand{\textsmallbulletfactor}{0.825}
\newcommand{\textsmallbulletscale}{0.5}
\newcommand{\textsmallbullet}{%
  \begingroup
  \sbox0{\textbullet}%
  
  \dimen0=\dimexpr\ht0-\textsmallbulletfactor\wd0\relax
  \dimen2=\dimexpr(\ht0+\dimen0)/2\relax 
  \sbox0{\scalebox{\textsmallbulletscale}{\raisebox{-\dimen0}{\textbullet}}}%
  \raisebox{\dimexpr\dimen2-0.5\ht0\relax}{\usebox0}%
  \endgroup
}

\title{Towards Virtual Networking for a Time-Triggered Hypervisor}

\author{ \href{https://orcid.org/0000-0002-0305-7139}{\includegraphics[scale=0.06]{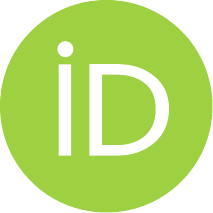}\hspace{1mm}Jan Ruh} \\
	NXP Semiconductors,\\
    Vienna, Austria\\
	\texttt{jan.ruh@nxp.com} \\
	\And
	\href{https://orcid.org/0000-0003-0116-9265}{\includegraphics[scale=0.06]{orcid.pdf}\hspace{1mm}Silviu S. Craciunas} \\
	NXP Semiconductors,\\
    Vienna, Austria\\
	\texttt{silviu.craciunas@nxp.com} \\
    \And
	Jan Korecky \\
	TTControl,\\
    Vienna, Austria
}

\renewcommand{\shorttitle}{Towards Virtual Networking for a Time-Triggered Hypervisor}

\hypersetup{
pdftitle={Towards Virtual Networking for a Time-Triggered Hypervisor},
pdfauthor={Jan Ruh, Silviu S. Craciunas, Jan Korecky},
pdfkeywords={Virtual Networking, Hypervisor, Real-time, Time-Triggered.},
}

\begin{document}
\maketitle

\begingroup
\renewcommand{\thefootnote}{}
\footnotetext{RTNS 2025, November 5--7, 2025, Pisa, Italy.
\textcopyright{} 2025 Copyright held by the owner/author(s).
Publication rights licensed to Springer.
This is the author's version of the work. It is posted here for
your personal use. Not for redistribution.
The definitive Version of Record was published in
\textit{The 33rd International Conference on Real-Time Networks and
Systems (RTNS 2025)}, November 5--7, 2025, Pisa, Italy.}
\endgroup

\newtheorem{constraint}{Constraint}
\theoremstyle{lemma}

\begin{abstract}
	Virtualization enables mixed-criticality applications to run on shared hardware by isolating them within separate virtual machines (VMs). However, achieving end-to-end timing guarantees for critical tasks hosted on virtual CPUs (VCPUs) without relying on static partitioning is hard. A possible solution is a time-triggered hypervisor that dispatches VMs' VCPUs to processor cores according to a predetermined global schedule. A global schedule includes the computation, the network, and the virtualization layer to provide end-to-end timing guarantees across hosts and VMs during design time. In this work, we propose a VirtIO-based virtual network that realizes IEEE 802.1Qbv quasi-compliant inter-VM communication, providing predictable timing for critical messages exchanged between time-triggered VMs. We present our VirtIO-based virtual network architecture and derive network-level scheduling constraints that emerge from the properties of the proposed solution. We implement the proposed design in the ACRN hypervisor and evaluate the end-to-end latency of critical traffic in a loaded network, demonstrating that our time-triggered VirtIO-net prototype provides predictable end-to-end latency under load.
\end{abstract}


\section{Introduction}
\label{sec:introduction}

Modern cyber-physical systems (CPS) are becoming increasingly powerful, featuring enhanced processing capabilities, a greater number of CPU cores, and expanded memory. This evolution allows for a broader range of functionalities to execute on a single device by running multiple virtual machines (VMs). Typically, we require CPS to deliver both real-time functionalities and high-level user-facing applications~\cite{Heiser11}. Virtualization thrives in edge computing for industrial automation, where data processing takes place at the network's edge in proximity to machinery rather than in the cloud. Here, edge computing devices run hard real-time control systems alongside soft real-time or best-effort data-driven applications~\cite{Breivold15}. Moreover, the automotive industry shifts away from using dedicated electronic control units (ECU) per function to utilizing computationally more powerful multicore multi-SoC systems that support both infotainment applications and conventional automotive control functions~\cite{mckinsey}, \cite{niedrist}. Additionally, high-level functionalities in CPS often require feature-rich general-purpose operating systems, such as GNU/Linux derivatives, while safety-critical and real-time functionalities rely real-time operating systems~\cite{Heiser11}.

In full system virtualization, a hypervisor creates a hardware-software duplicate of a real machine allocating resources to the resulting VMs~\cite{Popek74}. There are two main approaches to tackle real-time requirements in VMs: consolidating VMs' resource requirements or statically partitioning hardware resources. The former allocates virtual CPUs (vCPUs) to VMs and the hypervisor dispatches VCPUs on processor cores according to a runtime scheduling policy. For example, RT-XEN~\cite{Xi2011}, \cite{Xi2014}, implements a real-time deferrable server and uses compositional schedulability analysis~\cite{DBLP:journals/rts/EaswaranSL09} to dimension VCPU budgets and periods for hosted (soft) real-time tasks. Conversely, resource partitioning~\cite{DBLP:journals/corr/RamsauerKLM17,DBLP:conf/vee/LiXRD19,DBLP:conf/hipeac/MartinsTSBP20} statically assigns hardware (HW) resources, such as processor cores, memory, and I/O devices, to VMs to achieve strict spatial isolation, temporal isolation, and freedom from interference. Static partitioning trades timing predictability for scalability and efficiency since the available HW resources, such as cores or network interface cards (NICs) limits the number of VMs~\cite{Heiser11}. A \emph{time-triggered hypervisor}~\cite{Ruh21,Ruh21b,DBLP:conf/rtns/RuhC24} is a compromise that dispatches VCPUs on processor cores according to a precomputed static schedule table to enable strict temporal isolation with design-time end-to-end timing guarantees~\cite{DBLP:conf/rtns/RuhC24} while static memory partitions enforce spatial isolation. Event-driven VCPUs can utilize idle slots within the time-triggered schedule~\cite{FinziRTAS24} offering resource efficiency.

In many applications, VMs need to communicate with each other, on the same or different hosts, and with additional requirements on the maximum end-to-end latency of critical data transmission. For inter-VM communication on the same host, shared memory offers the least computational overhead, and the inter-VM shared memory~(IVSHMEM) device~\cite{Macdonell2011} supported in open-source hypervisors such as QEMU/KVM, ACRN~\cite{DBLP:conf/vee/LiXRD19}, and Jailhouse~\cite{DBLP:journals/corr/RamsauerKLM17}, can achieve high data throughput and reduced latency depending on the communication partners' timing characteristics. However, the peer-to-peer nature hinders applying standard protocols and inter-VM communication across hosts. In comparison, sharing NICs via virtual networks enables seamless inter-VM communication utilizing existing protocols and communication middleware across both VM and bare-metal deployments. Yet multiple VMs transmitting data on a virtual network entails unpredictable delays, which are unacceptable in real-time systems since they result in deadline misses and system failures. Unpredictable delays, trace back to runtime scheduling decisions in the virtual network backend that governs packet transmissions between VMs virtual NICs and external physical NICs so that it is hard to meet timing requirements in virtual networks.

In this paper we propose a time-triggered inter-VM communication architecture for a time-triggered hypervisor (such as the one proposed in~\cite{Ruh21,Ruh21b,DBLP:conf/rtns/RuhC24}) to enable predictable communication latencies between VMs on the same and different hosts. We derive network-level scheduling constraints that arise from the properties of the proposed design. We further describe the implementation of our proposed solution in the open-source hypervisor ACRN~\cite{DBLP:conf/vee/LiXRD19,TLF21} using a virtual network based on the VirtIO framework~\cite{Russell08}. Lastly, we evaluate the design on real industrial edge devices by testing end-to-end latencies of transferred packets under load.

We survey related work in Sec.~\ref{sec:rel_work}. Next, we present the system model (Sec.~\ref{sec:system_model}), the correctness constraints for time-triggered schedules (Sec.~\ref{sec:correctness_constraints}), and the implementation for the ACRN hypervisor (Sec.~\ref{sec:implementation}). In Sec.~\ref{sec:evaluation}, we outline the test setup and cases for evaluation, discuss results, and conclude in Sec.~\ref{sec:conclusion}.

\section{Related work}
\label{sec:rel_work}

This work examines time-triggered inter-VM communication based on the IEEE 802.1Qbv TSN amendment~\cite{Qbv}, aiming to extend time-triggered operations of a hypervisor to inter-VM communication, as suggested by~\cite{Ruh21,Ruh21b,DBLP:conf/rtns/RuhC24}. We review related work on time-triggered scheduling implementations at the user, operating system, and hypervisor levels, along with TSN features in Linux and inter-process and inter-VM communication methods.

\sloppypar
MotionWise~\cite{motionwise} implements a user-space TT scheduler that provides deterministic execution of automotive applications~\cite{motionwise_article}. Gala~et~al.~\cite{Gala2021} successfully applied time-triggered VCPU scheduling to Linux Kernel Virtual Machine~(KVM) in a cloud-computing environment to enforce temporal isolation and memory bandwidth limits. Karachatzis~et~al.~\cite{Karachatzis23} present a kernel-level TT scheduler for Linux and compare it to user space scheduling in MotionWise. Furthermore, Gala~et~al.~\cite{gala2023} present time-triggered Linux kernel-level scheduling with an acceptance test for event-triggered tasks. GNU/Linux implements several TSN features~\cite{TSN_docs}. For example, the traffic control utility (\lstinline|tc|)~\cite{tc} interfaces with the kernel's queueing disciplines~(\lstinline|qdiscs|) including the TAPRIO \lstinline|qdisc|~\cite{qdiscs}, providing scheduled traffic according to IEEE\,802.1Qbv. 

A common approach to inter-process and inter-VM communication is shared memory. The inter-VM shared memory~(IVSHMEM) device~\cite{Macdonell2011}, supported in open-source hypervisors such as QEMU/KVM, ACRN~\cite{DBLP:conf/vee/LiXRD19}, and Jailhouse~\cite{DBLP:journals/corr/RamsauerKLM17} allows sharing share memory with co-hosted VMs through a virtual PCI device. In recent versions, Jailhouse implements an extended IVSHMEM specification v2.0, enabling connections between up to $65536$ peers, interrupt-based event signaling, and the option to use an IVSHMEM network driver that exposes a virtual Ethernet device~\cite{DBLP:journals/rts/FerraroPGGBRBS23}. Similar virtual PCI shared memory devices can be found in commercial offerings such as QNX~\cite{qnx_shmem}. Alternatively, hypervisors that feature a Linux host, such as the KVM, or a Linux-based privileged VM, such as Xen~\cite{DBLP:conf/sosp/BarhamDFHHHN03} or ACRN~\cite{DBLP:conf/vee/LiXRD19}, often realize inter-VM communication as a virtual network with TAP devices attached to a Linux bridge appearing to the VMs as regular network interfaces. A memory region backs up each TAP device shared between the VM and the system hosting the Linux bridge. The de facto standard for virtual network devices in Linux-based virtual networks is VirtIO~\cite{Russell08}, which specifies how virtual device backends in the privileged system communicate data via \lstinline|virtqueues| with virtual device frontends in the VMs.

In~\cite{DBLP:conf/etfa/LeonardiBP20}, three alternative placements of a transmission (TX) scheduler for TSN in virtualized platforms, i.e., one per application, one per VM, or one in the hypervisor/a privileged context, are evaluated using simulation. However, they fail to acknowledge that relying on a TX scheduler in either the user space or the kernel space of VMs requires VMs to trust each other to adhere to the schedule. We argue that the TX scheduler must reside in a trusted component hindering the practicality of their solution. Schwäricke~et~al.~\cite{Schwäricke2021} propose a framework enabling predictable inter-VM communication in static partitioning hypervisors. They implemented a prototype in Jailhouse, delegating stream scheduling to a dedicated bare-metal broker cell that exploits the hardware platform's DMA engine for data transfer and utilizes established mechanisms, including cache-coloring~\cite{DBLP:conf/rtas/MancusoDBCCP13} and memory bandwidth regulation~\cite{DBLP:journals/tc/YunYPCS16} to isolate EDF-scheduled streams between VMs. VMs interface with the predictable inter-VM communication engine using standard VirtIO socket virtual devices. Li et al.~\cite{Li2023} propose a real-time communication framework utilizing an IVSHMEM memory region shared between multiple VMs. The shared memory region for each VM contains distinct TX and reception (RX) queues per priority, as well as a priority-based TX engine. We note that fixed priority-based stream scheduling is prone to starvation in the case of a faulty high-priority sender. Furthermore, the TX engines' execution context, i.e., whether it executes as a separate process within the VMs or the context of the transmitting application as part of the IVSHMEM device driver, or is it a host OS component, remains unclear despite its implications on expected throughput, temporal isolation, and security. Ferraro~et~al.~\cite{DBLP:journals/rts/FerraroPGGBRBS23} introduce a similar solution as part of their automotive time-sensitive architecture. Their prototypical implementation of an IVSHMEM-based virtual TSN switch~(VTSN) executes as a bare-metal application in a dedicated Jailhouse cell. It exhibits reduced round-trip time for varying packet sizes compared to a partitioned Xen virtual networking configuration. However, the current VTSN prototype does not actually support any TSN traffic shaping mechanisms. BiRtIO~\cite{DBLP:journals/access/PeixotoMCP24} is a VirtIO network implementation for the Bao static partitioning hypervisor. The authors report an average latency jitter of $243.96\,\mu s$ using multi-queue priority (MQPRIO) and credit-based shaper (CBS) \lstinline|qdiscs|.

\section{System Model}
\label{sec:system_model}
\begin{figure}[t!]
\centering
	\includegraphics[width=0.85\textwidth,trim={0.1cm 0.0cm 1cm 0.1cm},clip]{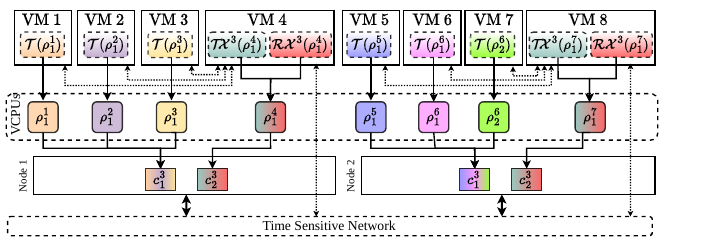}
	\caption{System model with a virtual network}
	\label{fig:system_model}
\end{figure}

We consider a distributed virtualized real-time system and illustrate an example with two nodes in Fig.~\ref{fig:system_model}. We model it as a directed graph $G(\mathcal{V},\mathcal{L})$ where the vertices $\mathcal{V}=\mathcal{V}^{S}\cup\Big(\bigcup_{\forall n\in\mathcal{V}^N}\mathcal{V}^n\Big)$ are switches $\mathcal{V}^S$ and VMs $\mathcal{V}^n$ hosted on virtualized end-systems $n\in\mathcal{V}^N$. The edges $\mathcal{L}\subset \mathcal{V}\times\mathcal{V}$ are either bi-directional, full-duplex physical links between vertices (VMs or switches) or bi-directional \emph{virtual links} between VMs hosted on the same end-system. A virtualized end-system is an SMP platform with a table-driven dispatcher in the hypervisor, the privileged VM, and user VMs. The hypervisor allocates VCPUs on physical cores (c.f. Fig.~\ref{fig:system_model}, VCPUs), while the privileged VM (c.f. Fig.~\ref{fig:system_model}, VMs~$4$~and~$8$) manages virtual links for its user VMs. User VMs (c.f. Fig.~\ref{fig:system_model}, VMs~$1$--$3$ and $5$--$7$) run periodic TT tasks based on a coordinated schedule. These VMs execute a set of dependent and independent TT tasks, with dependent tasks communicating through streams on the physical/virtual TSN backbone.
Each switch and end system's privileged VM executes the IEEE\,802.1AS clock synchronization protocol~\cite{802.1AS} to establish a global time base~\cite{DBLP:books/sp/KopetzS22} with an upper bounded precision $\Pi$, i.e., the maximum time difference between any two non-faulty clocks. Privileged VMs share the derived synchronized time with co-located user VMs and the hypervisor via a shared memory-based dependent clock~\cite{Ruh21}. The \emph{macrotick} (slot length) divides the global time in the guest OS and hypervisor of a virtualized end system into equal slices, representing the smallest granularity for scheduling decisions~\cite{DBLP:conf/icdcs/Kopetz92}, \cite{DBLP:journals/rts/IsovicF09}. To simplify the already complex correctness constraints, we will assume a timeline represented by the natural numbers, i.e., a macrotick of $1$ on all nodes. Generalizing the correctness constraints to different macroticks is trivial (e.g.,~\cite{DBLP:conf/rtns/RuhC24}). A static schedule of a node's core or a link's egress queue is a set of slots that repeats with a period called the hyperperiod~$T_{HP}$.

For completeness, we recapitulate the physical network and communication model, the task model, and the system virtualization model from~\cite{DBLP:conf/rtns/RuhC24} in Sec.~\ref{ssec:physical_network_and_communication_model}. We introduce our virtual network model in Sec.~\ref{ssec:virtual_network_model}.

\subsection{Communication, Task, and Virtualization Model}
\label{ssec:physical_network_and_communication_model}
The physical network is a time-sensitive network~(TSN)~\cite{TSN} featuring IEEE\,802.1As-2020~\cite{802.1AS}, gPTP clock synchronization and IEEE\,802.1Qbv~\cite{802.1Qbv} scheduled traffic. IEEE 802.1Qbv \cite{802.1Qbv} introduces a time-aware gate mechanism that manages the opening and closing of egress ports, thereby controlling the forwarding of frames based on a cyclic schedule table, referred to as a Gate Control List (GCL) \cite{DBLP:conf/rtns/CraciunasOCS16}. A NIC's local clock governs its operation~\cite{DBLP:conf/wfcs/CraciunasO21}. A stream $s_i:=\{(a,\nu_1),(\nu_1,\nu_2),...,(\nu_{m-1},\nu_m),(\nu_m,b)\}$ between a sender $a\in\mathcal{V}$ and a receiver $b\in\mathcal{V}$ on a predefined route via intermediate nodes $\nu_1,...,\nu_m\in N$ is a communication requirement with period $T_i$. Each stream $s_i\in\mathcal{S}$ periodically releases stream jobs that must be transmitted within their respective period instances. The communication model is not strictly periodic, i.e., each stream job can be transmitted at a different time in each period instance. Let $\mathcal{J}_i^{xy}$ be the finite ordered set of jobs that stream $s_i\in\mathcal{S}$ releases on a link $(x,y)\in s_i$ until the hyperperiod. Each stream $s_i\in\mathcal{S}$ communicates a payload of size $S_i$. When communicating via physical links, the maximum transmission unit~($MTU$) limits the number of bytes in a frame so that each stream job $J^{xy}_{ij}\in\mathcal{J}_i^{xy}$ consists of a list of \emph{frames} $\mathcal{F}^{xy}_{ij}$ with $|\mathcal{F}_{ij}^{xy}|=\ceil{\frac{S_i}{MTU}}$ since a stream's total payload can exceed the $MTU$. Each frame $f_{ijk}^{xy} \in \mathcal{F}_{ij}^{xy}$ has a transmission duration $f_{ijk}^{xy}.L$. When transmitted over a physical link~$(x,y)\in s_i$ with a given speed $s^{x,y}$ in bits/second, the transmission duration is $f_{ijk}^{xy}.L = \frac{S_{ijk} \cdot 10^9 \cdot 8}{s^{xy}},$ where $S_{ijk}$ is the size of the frame $f_{ijk}^{xy}$ in bytes. The transmission time of a frame $f_{ijk}^{xy} \in \mathcal{F}_{ij}^{xy}$ on link~$(x,y)\in s_i$ relative to the period is $f_{ijk}^{xy}.\phi$. The absolute scheduling offset of a frame $f_{ijk}^{xy}$ in period instance $j$ is $f_{ijk}^{xy}.\phi + j\cdot T_i$.  

A TT task $\tau_i$ is a tuple $(R_i,D_i,C_i,T_i,A_i)$ where $R_i$ is the release time, $D_i$ is the relative deadline, $C_i$ is the WCET, $T_i$ is the period, and $A_i$ is its resource affinity. The set of all periodic RT tasks is~$\mathcal{T}$. A context switch between tasks on a core $c$ induces an overhead $\delta_{\tau}(c)$. As for the streams, we assume a non-strictly periodic model, i.e., each task job can execute at a different time in each period instance. A task $\tau_i$ releases a finite ordered set of jobs $\mathcal{J}_i$ until the hyperperiod. Jobs are preemptible, so each job $J_{ij} \in \mathcal{J}_i$ comprises a finite list of \emph{job segments} $\mathcal{X}_{ij}$. A period-relative offset $x_{ijk}.\phi$ and length~$x_{ijk}.L$ define a job segment $x_{ijk} \in \mathcal{X}_{ij}$. A segment's absolute scheduling offset in period instance $j$ is $x_{ijk}.\phi + j T_i$. A pair of TT tasks communicate with each other via a TSN stream. Let $\tau_i\xrightarrow{s_k}\tau_j$ indicate a data dependency between two tasks $\tau_i,\tau_j\in\mathcal{T}$, i.e., $\tau_j$ receives input data from $\tau_i$ via stream $s_k\in\mathcal{S}$. It is $E(\tau_i\xrightarrow{s_k}\tau_j)$ the maximum end-to-end latency of a data dependency.

A virtualized end system $n\in \mathcal{V}^N$ is an SMP system, and $\mathcal{C}^n$ denotes its set of cores that share a main memory. Each virtualized end system hosts multiple TT VMs that share the physical compute and memory resources. TT VMs can have access to a dedicated network interface card~(NIC) connected to the TSN backbone or to a virtual NIC connected to a software bridge in a \emph{privileged} VM as illustrated in Fig.,~\ref{fig:system_model}. A TT VM $j\in \mathcal{V}^n$ dispatches TT tasks on a set of VCPUs. Each VCPU is assigned to run on exactly one physical core, with no migration allowed at runtime. Let $\mathcal{P}(c)$ be the set of all VCPUs running on a core $c\in\mathcal{C}^n$. TT VCPUs are preemptible; thus, a TT VCPU $\rho_i\in\mathcal{P}(c)$ comprises a set of segments $\mathcal{Z}_i$. An offset $z_{ij}.\phi$ and a length $z_{ij}.L$ define a VCPU segment $z_{ij} \in \mathcal{Z}_i$. VCPU segments do not have a period, and their offset is relative to the hyperperiod. Switching between VCPUs on a core $c$ introduces a context switch overhead $\delta_{\rho}(c)$. Let $\mathcal{T}(\rho)$ be the set of TT tasks allocated to a VM's VCPU $\rho$.

\subsection{Virtual Network Model}
\label{ssec:virtual_network_model}
In this paper, we propose a time-triggered IEEE\,802.1Qbv quasi-compliant inter-VM communication based on the VirtIO specification~\cite{virtio-v1.2}. On each virtualized end system $n\in\mathcal{V}^N$, we assume the existence of a privileged VM $p\in\mathcal{V}^n$, hosting a software bridge. The software bridge forwards Ethernet frames between co-located user VMs and to the TSN backbone connected to the privileged VM through a passthrough NIC. There exists a shared memory region between each user and privileged VM, which in turn, contains $q$ egress and ingress queues. Towards a user VM $v\in\mathcal{V}^n$, its egress and ingress queues appear as a regular NIC, whereas the privileged VM's OS kernel abstracts each shared memory region as a kernel virtual network interface (in UNIX-like OS commonly known as a TAP interface). The privileged VM serves a VM $v$'s egress and ingress queues with dedicated transmission (TX) and reception (RX) tasks $\mathcal{TX}^{v}(\rho_p):=\{\tau^{tx}_{v,1},...,\tau^{tx}_{v,q}\}$ and $\mathcal{RX}^{v}(\rho_p):=\{\tau^{rx}_{v,1},...,\tau^{rx}_{v,q}\}$ on a virtual CPU $\rho_p$. We refer to the shared memory region between a user VM $u\in\mathcal{V}^n\setminus\{p\}$ and the privileged VM $p$, the egress and ingress queues it contains, and the TX and RX tasks serving them as a virtual link $(v,p)\in\mathcal{L}$.

Notably, we consider bidirectional links, i.e., $(v,p)=(p,v)\in\mathcal{L}$, however, if we write a virtual link as $(v,p)$, it implies that the privileged VM must dispatch $v$'s TX task $\tau^{tx}_{v,i}\in\mathcal{TX}^{v}(\rho_p)$ to \emph{read} sent Ethernet frames from its corresponding egress queue $i$ and let the software bridge determine their destination. Accordingly, if we write a virtual link as $(p,v)$, it implies that $v$'s RX task $\tau^{rx}_{v,i}\in\mathcal{RX}^{v}(\rho_p)$ \emph{writes} the Ethernet frames forwarded by the software bridge to the corresponding ingress queue $i$. In contrast to physical networks, where a frame's transmission duration $f.L$ depends on the link speed, in a virtual network, the worst-case memory read and write rates $rd^{wc}$ and $wr^{wc}$ of the virtualized end system hosting a virtual link capture the transmission duration depending on whether the virtual link is sending or receiving. If a virtual link $(v,p)$ is sending, i.e. its TX task $\tau^{tx}_{v,i}$ reads data from the egress queue $i$, the transmission duration is $f^{vp}_{lkm}.L=\frac{S_{lkm}}{rd^{wc}}=f^{vp}_{lkm}.L^{rd}$. If a virtual link $(p,v)$ is receiving, i.e., its RX task $\tau^{rx}_{v,i}$ writes data to the ingress queue $i$, the transmission duration is $f^{pv}_{lkm}.L=\frac{S_{lkm}}{wr^{wc}}=f^{pv}_{lkm}.L^{wr}$.

We denote the set of all TX and RX tasks assigned to a VCPU $\rho_{p}\in\mathcal{P}(c)$ of a privileged VM $p$ with $\mathcal{TX}^n(\rho_p):=\bigcup_{\forall v\in\mathcal{V}^n}\mathcal{TX}^v(\rho_p)$ and $\mathcal{RX}^n(\rho_p):=\bigcup_{\forall v\in\mathcal{V}^n}\mathcal{RX}^v(\rho_p)$. Furthermore, we write $\mathcal{S}^{tx}_{v,i}$, $\mathcal{S}^{rx}_{v,i}$, $\mathcal{J}^{tx}_{v,i}$, and $\mathcal{J}^{rx}_{v,i}$ for the sets of all streams and stream jobs send and received by a VM $v\in\mathcal{V}^n$ that map to egress or ingress queue $1\leq i\leq q$, respectively.

We make no assumptions on the number of VCPUs available to a privileged VM to host virtual links' TX and RX tasks, nor its VCPUs' execution mode, i.e., VCPUs hosting virtual links may be pinned to a processor core or time-triggered. While a TX task $\tau^{tx}_{v,i}$ or RX task $\tau^{rx}_{v,i}$ is active on its VCPU $\rho_p$, it forwards or receives queued frames to or from the respective egress or ingress queues of its virtual links. Accordingly, dispatching an egress queue $i$'s TX task $\tau^{tx}_{v,i}$ corresponds to setting the state of a 802.1Qbv time-aware gate on VM $v$'s virtual NIC's $i$-th egress queue to open. It becomes apparent that the timing, i.e., latency and jitter, of data transmission via virtual links depends on scheduling TX and RX tasks' jobs on the VCPU $\rho_p$. We assume that TX and RX tasks only release new jobs if packets are waiting in their respective egress and ingress queues so that the task model for scheduling TX and RX tasks follows the characteristics of the TSN streams generating data.

Consequently, a VM $v\in\mathcal{V}^n$ communicating Qbv streams over a virtual link will periodically place packets in a TT egress queue $i$ or retrieve packets from a TT ingress queue $i$. Therefore, TX and RX task $\tau^{tx}_{v,i}\in\mathcal{TX}^v(\rho_p)$ and $\tau^{rx}_{v,i}\in\mathcal{TX}^v(\rho_p)$ are TT tasks $(R_i^v,D_i^v,C_i^v,T_i^v,A_i^v)$ with release time $R_i^v=0$, a period $T_i^v=\lcm\{T_x|\forall s_x\in\mathcal{S}^{tx}_{v,i}\}$, a deadline $D_i^v=T_i^v$, and an affinity $A_i^v:={\rho_p}$. The WCET $C_i^v$ of a TX task is the sum of its frames' worst-case read durations and TX delay $\delta^{tx}$ times the number of processed stream jobs $\mathcal{J}^{tx}_{v,i}$:
\begin{align}
C_i^v=\delta^{tx}\cdot |\mathcal{J}^{tx}_{vi}| +\sum_{\forall J^{tx}_{vij}\in\mathcal{J}^{tx}_{vi}}\Big(\sum_{f^{v}_{ijk}\in\mathcal{F}^{tx}_{vij}}f^{v}_{ijk}.L^{rd}\Big)\nonumber
\end{align}
A RX task's WCET $C_i^v$ is the sum of its frames' worst-case write durations and a \emph{reception delay} $\delta^{rx}$ times the number of processed stream jobs $\mathcal{J}^{rx}_{v,i}$:
\begin{align}
C_i^v=\delta^{rx}\cdot |\mathcal{J}^{rx}_{vi}| + \sum_{\forall J^{rx}_{vij}\in\mathcal{J}^{rx}_{vi}}\Big(\sum_{f^{v}_{ijk}\in\mathcal{F}^{rx}_{vij}}f^{v}_{ijk}.L^{wr}\Big)\nonumber
\end{align}
If we are not explicitly referencing a specific VM's TX or RX task, we drop the VM superscripts for readability. A stream $s_x\in\mathcal{S}$ features at most two virtual links, one between the sender and the privileged VM and one between the privileged VM and the receiver.

Given the sets of periodic TX and RX tasks $\mathcal{TX}(\rho_p)$ and $\mathcal{RX}(\rho_p)$ we can apply the TT task model. Up to the hyperperiod $T_{HP}$, for all TT queues $i\in\{1,...,m\}$, TX and RX tasks $\tau^{tx}_{i}\in\mathcal{TX}(\rho_p)$ and $\tau^{rx}_{i}\in\mathcal{RX}(\rho_p)$ release a finite number of jobs denoted by $\mathcal{J}^{tx}_i$ and $\mathcal{J}^{rx}_i$. We assume that TX and RX tasks' jobs are preemptable, i.e., each job $J^{tx}_{ij}\in\mathcal{J}^{tx}_i$ and $J^{rx}_{ij}\in\mathcal{J}^{rx}_i$ gives rise to finite lists of job segments $\mathcal{X}^{tx}_{ij}$ and $\mathcal{X}^{rx}_{ij}$. We define a TX or RX task job segment $x_{ijk}$ by its period-relative offset $x_{ijk}.\phi$ and its length $x_{ijk}.L$.

\section{Correctness Constraints}
\label{sec:correctness_constraints}
Previous work~\cite{DBLP:conf/rtns/RuhC24} proposed correctness constraints for TT tasks with end-to-end latency on TT VCPUs in SMP platforms using a TSN backbone. However, it assumed each TT VM had a passthrough NIC, which isn’t usually the case. We address this by introducing new correctness constraints that enable IEEE\,802.1Qbv inter-VM and external communication through virtual NICs, as detailed in Sec.~\ref{ssec:virtual_network_model}. The scheduling problem, i.e., if for a given set of TT tasks hosted on a set of TT VCPUs executing on cores across several virtualized end systems communicating via Qbv streams over a mixed physical/virtual TSN backbone there exist(s) correct solution(s) that fulfill(s) the correctness constraints, is NP-complete. A solution of the scheduling problem is a set of static schedule tables specifying task and VCPU segment offsets, lengths, and frame offsets fulfilling all correctness constraints. We differentiate between checking a given solution's correctness, which can be done in polynomial time, and finding a solution for which (if $P\neq NP$) no polynomial time algorithm exists. For now, we are only interested in the correctness of an existing solution.

Akin to the introduced system model, in Sec.~\ref{ssec:task_correctness}--\ref{ssec:sys_virt_constr} we briefly recapitulate the most important correctness constraints presented in~\cite{DBLP:conf/rtns/RuhC24}, and introduce new constraints (Strict Source and Destination Task Alignment Constraints) required by our generalization of the model in~\cite{DBLP:conf/rtns/RuhC24}. Then, in Sec.~\ref{ssec:virtual_network_constraints}, we introduce the correctness constraints for the proposed virtual network model that extend the existing network constraints~\cite{DBLP:conf/rtns/RuhC24} enabling strict, predictable end-to-end latency when communicating in a mixed physical/virtual Qbv network.

\subsection{Task Correctness Constraints}
\label{ssec:task_correctness}

Job segments must be limited by release, deadline, and WCET of their tasks.
\begin{constraint}[Release Time and Deadline]
	\label{const:release_time_deadline_wcet}
	Given the virtualized distributed real-time system defined by the graph $G(\mathcal{V},\mathcal{L})$, we have:
	\begin{align}
	& \forall n\in \mathcal{V}, \forall c \in \mathcal{C}^n, \forall \rho \in \mathcal{P}(c), \forall\tau_i\in\mathcal{T}(\rho),\forall J_{ij}\in\mathcal{J}_i,\forall x_{ijk}\in\mathcal{X}_{ij}:\nonumber\\
	& (x_{ijk}.\phi \geq R_i) \wedge (x_{ijk}.\phi+x_{ijk}.L\leq D_i)\nonumber
	\end{align}
\end{constraint}

The length of each task segment should be no shorter than the overhead of context switching to ensure job progression. Additionally, the total size of a job's segments must equal or exceed the task's Worst Case Execution Time (WCET) plus the context-switching overhead for those segments.
\begin{constraint}[Job Segment Size Constraint]
	\label{const:job_segment_size_constraint}
	Given the virtualized distributed real-time system defined by the graph $G(\mathcal{V},\mathcal{L})$, we have: 
	\begin{align}
	& \forall n\in \mathcal{V}, \forall c \in \mathcal{C}^n, \forall \rho \in \mathcal{P}(c), \nonumber \forall \tau_i\in \mathcal{T}(\rho) ,\forall J_{i,j}\in\mathcal{J}_i:\nonumber\\
	& (\forall x_{i,j,k}\in\mathcal{X}_{ij}: x_{i,j,k}.L\ge \delta_{\tau}(c)) \wedge  \left(\sum_{\forall x_{i,j,k}\in\mathcal{X}_{ij}}x_{i,j,k}.L \geq C_i + |\mathcal{X}_{i,j}|\cdot \delta_{\tau}(c) \right)\nonumber
	\end{align}
\end{constraint}

For any pair of tasks assigned to the same resource, their job segments' execution intervals cannot overlap.
\begin{constraint}[Task Overlapping Constraint]
	\label{const:task_overlapping}
	Given the virtualized distributed real-time system defined by the graph $G(\mathcal{V},\mathcal{L})$, we have:
	\begin{align}
	& \forall n\in \mathcal{V}, \forall c \in \mathcal{C}^n, \forall \rho_1, \rho_2 \in \mathcal{P}(c), \forall \tau_i \in \mathcal{T}(\rho_1), \forall \tau_l \in \mathcal{T}(\rho_2), \forall J_{ij}\in\mathcal{J}_i, \forall J_{lm}\in\mathcal{J}_l, \nonumber \\
    &\forall x_{ijk}\in\mathcal{X}_{ij}, \forall x_{lmn}\in\mathcal{X}_{lm}, x_{ijk} \neq x_{lmn}: (x_{ijk}.\phi + j\cdot T_i \geq \nonumber \\
    & x_{lmn}.\phi+ m\cdot T_l +x_{lmn}.L)\lor (x_{lmn}.\phi + m\cdot T_l\geq x_{ijk}.\phi + j\cdot T_i + x_{ijk}.L)\nonumber
	\end{align}
\end{constraint}

\subsection{End-to-End Correctness Constraints}
The end-to-end latency constraint imposes a maximum delay between the start of a sender task and the completion time of a receiver task in a user VM. Virtualized end systems and switches within a TSN network $G(\mathcal{V},\mathcal{L})$ share a common global time established by the IEEE\,802.1AS clock synchronization protocol and characterized by its precision $\Pi$. The time difference between a receiving task job completing its execution and a sending task job starting execution is less than or equal to their streams' end-to-end latency minus the precision $\Pi$. 

\begin{constraint}[End-to-End Constraint]
    \label{const:e2e_dependency}
    Given the virtualized distributed real-time system defined by the graph $G(\mathcal{V},\mathcal{L})$, we have: 
    \begin{align}
        & \forall a, b\in \mathcal{V}, \forall c_a \in \mathcal{C}^{a}, \forall c_b \in \mathcal{C}^{b}, \forall \rho_a \in \mathcal{P}(c_a), \forall \rho_b \in \mathcal{P}(c_b), \nonumber \\
        & \forall\tau_i \in\mathcal{T}(\rho_a), \tau_l \in\mathcal{T} (\rho_b) \text{ s.t. } \tau_i\xrightarrow{s_h}\tau_l, \forall J_{ij}\in\mathcal{J}_i, \forall J_{lm}\in\mathcal{J}_l: 
         \nonumber\\
        &\max_{\forall x_{lmn}\in\mathcal{X}_{lm}}\{x_{lmn}.\phi  + x_{lmn}.L\} -  \min_{\forall x_{ijk}\in\mathcal{X}_{ij}} \{x_{ijk}.\phi \} \leq E(\tau_i\xrightarrow{s_h}\tau_j) - \Pi \nonumber
    \end{align}
\end{constraint}

Furthermore, he transmission schedule of the stream {$s_k$} aligns along its route to the sender and receiver task schedule. The propagation delay $\delta^{xy}$ is the time it takes a frame to travel on a \emph{physical} link from the source to the destination.
\begin{constraint}[Task Alignment Constraint]
    \label{const:task_alignment_constraint}
    Given the set of streams $\mathcal{S}$ and the virtualized distributed real-time system defined by the graph $G(\mathcal{V},\mathcal{L})$, we have: 
    \begin{align}
        & \forall a, b\in \mathcal{V}, \forall c_a \in \mathcal{C}^{a}, \forall c_b \in \mathcal{C}^{b}, \forall \rho_a \in \mathcal{P}(c_a), \forall \rho_b \in \mathcal{P}(c_b), \nonumber \\
        & \forall\tau_i \in\mathcal{T}(\rho_a), \tau_j \in\mathcal{T} (\rho_b) \text{ s.t. } \tau_i\xrightarrow{s_h}\tau_j, \forall J_{ik}\in\mathcal{J}_i, \forall J_{jk}\in\mathcal{J}_j: \nonumber\\
        &\left ( \max_{\forall x_{ikl}\in\mathcal{X}_{ik}} \{x_{ikl}.\phi + x_{ikl}.L\} \leq \min_{\forall f_{hkl}^{ax}\in \mathcal{F}_{hk}^{ax}}\{f_{hkl}^{ax}.\phi \} \right ) \wedge \nonumber \\ 
        &\left( \min_{\forall x_{jkl}\in\mathcal{X}_{jk}} \{x_{jkl}.\phi\} \geq \max_{\forall f_{hkl}^{yb}\in \mathcal{F}_{hk}^{yb}} \{f_{hkl}^{yb}.\phi  + f_{hkl}^{yb}.L\} + \delta^{yb} + \Pi \right)\nonumber
    \end{align}
    
\end{constraint}
When a stream's source node is a VM, the VM's network stack buffers frames but its VCPU may not be running when the frames' TX or RX times arrive. Hence, in addition to the constraints from~\cite{DBLP:conf/rtns/RuhC24}, we introduce strict source and destination task alignment constraints. The strict source task alignment constraint ensures that the source task job's last segment and the transmission of the stream job's last frame complete simultaneously. The destination task alignment constraint ensures that a destination task job's first segment is dispatched when the TX of a stream job's first frame completes. Together with Constraint~\ref{const:virtual_cpu_wcet}, these new alignment constraints implicitly ensure that the source/destination task job's VCPU is running when the NIC transmits/receives a stream job's frames. We presume that the source/destination task's WCET includes the worst-case transmission/reception overhead induced by the network stack.
\begin{constraint}[Strict Source Task Alignment Constraint]
	\label{const:task_strict_snd_alignment_constraint}
	Given the set of streams $\mathcal{S}$ and the virtualized distributed real-time system $G(\mathcal{V},\mathcal{L})$, we have: 
	\begin{align}
	& \forall a\in\mathcal{V}, \forall b\in \mathcal{V}, \forall c_a \in \mathcal{C}^{a}, \forall c_b \in \mathcal{C}^{b}, \forall \rho_a \in \mathcal{P}(c_a), \forall \rho_b \in \mathcal{P}(c_b),\nonumber\\
	& \forall\tau_i \in\mathcal{T}(\rho_a), \tau_j \in\mathcal{T} (\rho_b) \text{ s.t. } \tau_i\xrightarrow{s_h}\tau_j, \forall J_{ik}\in\mathcal{J}_i, \forall J_{jk}\in\mathcal{J}_j: \nonumber\\
	&\max_{\forall x_{ikl}\in\mathcal{X}_{ik}} \{x_{ikl}.\phi  + x_{ikl}.L\} = \Bigg\lfloor\max_{\forall f_{hkl}^{ax}\in \mathcal{F}_{hk}^{ax}}\{f_{hkl}^{ax}.\phi+f_{hkl}^{ax}.L\}\Bigg\rfloor\nonumber
	\end{align}
\end{constraint}
For the destination task alignment constraint, if the last hop is a virtual link, we additionally need to account for the worst-case RX delay $\delta^{rx}$, the TX delay $\delta^{tx}$, and the context switch time $\delta_{\tau}(\rho_{p})$ to the virtual link's RX task job segment.
\begin{align}
	\delta^{rx}(v,u)=\begin{cases}
		\delta^{rx}+\delta^{tx}+\delta_{\tau}(\rho_{p}),&\text{ if }(v,u)\text{ is a virtual link}\\
		0,&\text{ otherwise}
	\end{cases}\nonumber
\end{align}
\begin{constraint}[Strict Destination Task Alignment Constraint]
	\label{const:task_strict_recv_alignment_constraint}
	Given the set of streams $\mathcal{S}$ and the virtualized distributed real-time system, we have: 
	\begin{align}
	& \forall a\in\mathcal{V}, \forall b\in \mathcal{V}, \forall c_a \in \mathcal{C}^{a}, \forall c_b \in \mathcal{C}^{b}, \forall \rho_a \in \mathcal{P}(c_a), \forall \rho_b \in \mathcal{P}(c_b),\nonumber\\
	& \forall\tau_i \in\mathcal{T}(\rho_a), \tau_j \in\mathcal{T} (\rho_b) \text{ s.t. } \tau_i\xrightarrow{s_h}\tau_j, \forall J_{ik}\in\mathcal{J}_i, \forall J_{jk}\in\mathcal{J}_j: \nonumber\\
	& \min_{\forall x_{jkl}\in\mathcal{X}_{jk}} \{x_{jkl}.\phi\} + \delta^{rx}(y,b) = \Bigg\lfloor\min_{\forall f_{hkl}^{yb}\in \mathcal{F}_{hk}^{yb}} \{f_{hkl}^{yb}.\phi + f_{hkl}^{yb}.L\}+\delta^{yb}+\Pi \Bigg\rfloor\nonumber
	\end{align}
\end{constraint}

\subsection{System Virtualization Constraints}
\label{ssec:sys_virt_constr}
Similar to Constraint~\ref{const:task_overlapping}, VCPU segments cannot overlap if on the same resource.
\begin{constraint}[Virtual CPU Overlapping]
	\label{const:virtual_cpu_overlap}
	Given the virtualized distributed real-time system defined by the graph $G(\mathcal{V},\mathcal{L})$, we have: 
	\begin{align}
	&\forall n\in \mathcal{V}, \forall c \in \mathcal{C}^n, \forall \rho_i, \rho_m \in\mathcal{P}(c), \forall z_{ij} \in \mathcal{Z}_i, \forall z_{mn} \in \mathcal{Z}_m, z_{ij} \neq z_{mn}: \nonumber \\
	& (z_{ij}.\phi \geq z_{mn}.\phi+z_{mn}.L)\lor (z_{mn}.\phi \geq z_{ij}.\phi+z_{ij}.L)\nonumber
	\end{align}
\end{constraint}

The VCPU segment size must be at least equal to the sum of all task job segments allocated to it, plus the overhead of VCPU context switching. This requirement involves a helper set $\mathcal{O}_{ij}$, which identifies the task job segments that completely overlap with the specific VCPU segment $z_{ij}$. For a VCPU segment $z_{ij}$ of $\rho_i$ assigned to core $c$, it can be expressed as: 
\begin{align}
& \mathcal{O}_{ij} = \{ x_{lmn} | \forall \tau_l \in \mathcal{T}(\rho_i), \forall J_{lm}\in\mathcal{J}_l, \forall x_{lmn}\in\mathcal{X}_{lm}: \nonumber \\  
& (x_{lmn}.\phi+ m\cdot T_l \ge z_{ij}.\phi) \wedge  (x_{lmn}.\phi + m\cdot T_l + x_{lmn}.L \le z_{ij}.\phi + z_{ij}.L) \} \nonumber
\end{align}

\begin{constraint}[Virtual CPU Size Constraint]
	\label{const:virtual_cpu_wcet}
	Given the virtualized distributed real-time system defined by the graph $G(\mathcal{V},\mathcal{L})$, we have: 
	\begin{align}
	&\forall n\in \mathcal{V}, \forall c \in \mathcal{C}^n, \forall \rho_i \in\mathcal{P}(c), \forall z_{ij} \in \mathcal{Z}_i:  z_{ij}.L \ge \delta_{\rho}(c) + \sum_{x_{lmn} \in \mathcal{O}_{ij}} x_{lmn}.L \nonumber
	\end{align}
\end{constraint}

All task segments that overlap with a given VCPU segment, i.e., are elements of the set $\mathcal{O}_{ij}$, must be assigned to the corresponding VCPU $\rho_i$. 
\begin{constraint}[Virtual CPU Assignment]
	\label{const:virtual_cpu_assignment}
	Given the virtualized distributed real-time system defined by the graph $G(\mathcal{V},\mathcal{L})$, we have: 
	\begin{align}
	&\forall n\in \mathcal{V}, \forall c \in \mathcal{C}^n, \forall \rho_i \in\mathcal{P}(c), \forall z_{ij} \in \mathcal{Z}_i, \forall x_{lmn} \in \mathcal{O}_{ij} : \tau_l \in \mathcal{T}(\rho_i) \nonumber
	\end{align}
\end{constraint}

\subsection{Virtual Network Constraints}
\label{ssec:virtual_network_constraints}
The correctness constraints for IEEE\,802.1Qbv from~\cite{DBLP:conf/rtns/RuhC24} also apply in this work. The following constraints specify correctness if a shared memory-based virtual network served by per-queue TX and RX tasks is present in the privileged VM of a virtualized end system. An 802.1Qbv stream's transmission on a virtual link differs from a deterministic Ethernet link in that a virtual link requires a TX or a RX task executing on a computational resource to forward data in an egress queue or to receive data on an ingress queue, respectively.

A VM $v\in\mathcal{V}^n$ sending a stream $s_k\in\mathcal{S}$ via a virtual link $(v,p)\in\mathcal{L}^n$, where $p\in\mathcal{V}^n$ is $n$'s privileged VM, amounts to writing the stream data to an egress queue $i$ and dispatching the corresponding TX task $\tau^{tx}_{v,i}\in\mathcal{TX}^v(\rho_{p})$. For released stream jobs $\mathcal{J}^{tx}_{v,i}$, the TX task $\tau^{tx}_{v,i}$ releases jobs $\mathcal{J}_{i}^{tx}$ to read the stream jobs' frames from $v$'s egress queue $i$ and determine their destination. Instead, if VM $v$ is receiving a stream $s_k$ on a virtual link $(p,v)\in \mathcal{L}^n$, it requires a RX task $\tau^{rx}_{v,i}\in\mathcal{RX}^v(\rho_{p})$ to write stream data to the associated ingress queue before its guest OS can read the frames. Thus, similar to outgoing streams, for incoming stream jobs $\mathcal{J}^{rx}_{v,i}$, the RX task $\tau^{rx}_{v,i}$ releases jobs $\mathcal{J}
_{i}^v$ to write the stream jobs' frames to $v$'s ingress queue $i$. As a result, a TX or RX task job $J_{ij}^v$ requires segments $\mathcal{X}^v_{ij}$ depending on the released stream job instances $J^{tx}_{kl}\in\mathcal{J}^{tx}_{v,i}$ or $J^{rx}_{kl}\in\mathcal{J}^{tx}_{v,i}$, i.e., for any stream job sent or received via a virtual link there must exist a TX or RX task job segment encapsulating and processing its frames. Accordingly, a TX/RX task job segment constraint must ensure that a TX or RX task job's segment size $x_{ijk}^v.L$ is greater or equal to the sum of all stream jobs' TX or RX delays, frames' transmission durations, and the task context switch time overhead $\delta_{\tau}(\rho_p)$ on a VCPU $\rho_p$. We express the constraint similar to the Virtual CPU Size Constraint~\ref{const:virtual_cpu_wcet} using a helper set. We define the helper set $\mathcal{R}_{ijk}^v$ containing all stream jobs a TX or RX task segment $x_{ijk}^v\in\mathcal{X}^v_{ij}$ serves:
\begin{align}
\mathcal{R}_{ijk}^v:=\{J^{tx}_{fg}|&\forall s_f\in\mathcal{S}^{tx}_{v,i},\forall J^{tx}_{fg}\in\mathcal{J}^{tx}_{v,i},\forall f_{fgh}\in \mathcal{F}^{tx}_{fg}: \nonumber \\
& j=\Big\lfloor\frac{g\cdot T_f}{T_i^v}\Big\rfloor \land (f_{fgh}.\phi+g\cdot T_f\geq x_{ijk}.\phi + j\cdot T_i) \nonumber\\
& \land (f_{fgh}.\phi+f_{fgh}.L+g\cdot T_f\leq x_{ijk}.\phi + x_{ijk}.L - \delta(x_{ijk}^v) + j\cdot T_i)\}\nonumber
\end{align}
The term $\delta(x_{ijk}^v)$ captures the TX delay $\delta^{tx}$, if $x_{ijk}^v$ is a TX task job segment and the RX delay $\delta^{rx}$, if $x_{ijk}^v$ is a RX task job segment:
\begin{align}
        \delta(x_{ijk}^v)=\begin{cases}\delta^{tx},&\text{ if } x_{ijk}^v\text{ is TX task job segment} \\
        -\delta^{rx},&\text{ if } x_{ijk}^v\text{ is RX task job segment}
    \end{cases}\nonumber
\end{align}
\begin{constraint}[TX Task Job Segment Size Constraint]
	\label{const:tx_task_constraint}
	Given the virtualized distributed real-time system defined by the graph $G(\mathcal{V},\mathcal{L})$, we have:
	\begin{align}
	& \forall n\in \mathcal{V}, \forall v\in\mathcal{V}^n, \forall c \in \mathcal{C}^n,\forall \rho_p \in\mathcal{P}(c),\forall \tau^{tx}_{v,i}\in\mathcal{TX}^v(\rho_p),\forall J^v_{ij}\in\mathcal{J}^v_{i},\forall x^v_{ijk}\in\mathcal{X}^v_{ij}\nonumber\\
    & \forall s_l\in\mathcal{S}^{tx}_{v,i},\forall J^{tx}_{lm}\in\mathcal{J}^{tx}_{lm}: x^v_{ijk}.L \geq \delta_{\tau}(\rho_p)+\sum_{\forall J^{tx}_{lm}\in \mathcal{R}_{ijk}^v}\Big(\delta^{tx}+\sum_{\forall f_{klm}\in J^{tx}_{lm}}f_{klm}.L\Big)\nonumber
	\end{align}
\end{constraint}
\begin{constraint}[RX Task Job Segment Size Constraint]
	\label{const:rx_task_constraint} 
	Given the virtualized distributed real-time system defined by the graph $G(\mathcal{V},\mathcal{L})$, we have:
	\begin{align}
	& \forall n\in \mathcal{V}, \forall v\in\mathcal{V}^n, \forall c \in \mathcal{C}^n,\forall \rho_p \in\mathcal{P}(c),\forall \tau^{tx}_{v,i}\in\mathcal{TX}^v(\rho_p),\forall J^v_{ij}\in\mathcal{J}^v_{i},\forall x^v_{ijk}\in\mathcal{X}^v_{ij}\nonumber\\
    & \forall s_l\in\mathcal{S}^{rx}_{v,i},\forall J^{rx}_{lm}\in\mathcal{J}^{tx}_{lm}:
	 x^v_{ijk}.L \geq \delta_{\tau}(\rho_p)+\sum_{\forall J^{rx}_{lm}\in \mathcal{R}_{ijk}^v}\Big(\delta^{rx}+\sum_{\forall f_{klm}\in J^{rx}_{lm}}f_{klm}.L\Big)\nonumber
	\end{align}
\end{constraint}

\section{Implementation}
\label{sec:implementation}

\subsection{SMT-EDF Schedule Co-Synthesis}
Encoding correctness constraints as linear integer arithmetic expressions ${(\mathbb{Z},+,\geq)}$ and using Satisfiability Modulo Theories~(SMT) to find valid variable assignments or a \emph{model} is an established method to schedule time-triggered systems~\cite{DBLP:conf/rtss/Steiner10,DBLP:conf/rtns/CraciunasO14,Craciunas16,DBLP:conf/rtns/CraciunasOCS16,DBLP:conf/rtns/RuhC24}. Incrementally adding assertions to, checking the SMT context, and backtracking if a SMT context is unsatisfiable has proven efficient for scheduling communication in TTEthernet~\cite{DBLP:conf/rtss/Steiner10}. However, when co-scheduling computational and communication resources of large systems, SMT-based scheduling with incremental backtracking exhibits long runtimes since the high utilization of computational compared to communication resources results in repeated backtracking~\cite{DBLP:conf/rtns/CraciunasO14}. The introduction of a system virtualization layer and VCPU constraints further multiplies interdependency, impeding the performance benefits gained by incremental backtracking. In~\cite{DBLP:conf/rtns/RuhC24}, the authors proposed identifying and jointly adding dependent VCPU, task, and stream constraints to the solver context to prevent conflicting segment offsets, segment sizes, and frame offsets of interdependent schedulable entities that otherwise would result in backtracking. Their SMT-based dependency-aware scheduler minimizes the number of required VCPU segments, hence virtualization overhead, with a reasonable runtime for small systems, yet it scales poorly with increasing system size. Therefore, they further implement a runtime-optimized scheduler similar to demand-based scheduling~\cite{DBLP:conf/rtns/CraciunasO14,Craciunas16} that utilizes the SMT scheduler to synthesize schedule tables for dependent tasks and their streams. Then, it simulates earliest deadline first (EDF) to schedule TT tasks without dependencies, so-called free tasks, and VCPUs via a heuristic. Their SMT-EDF scheduler effectively trades a decreased runtime for higher virtualization overhead~\cite{DBLP:conf/rtns/RuhC24}.

Here, we also implemented an SMT-EDF scheduler utilizing the Python API of Z3~\cite{DBLP:conf/tacas/MouraB08} to schedule dependent TT task jobs, their VCPUs, stream jobs, and TX and RX task jobs. We facilitate the resulting partial schedule tables to enforce busy intervals $\mathcal{I}^c$ on a processor core $c$ for the EDF simulation. Our SMT-EDF scheduler takes sets of TT tasks $\mathcal{T}$, TT TX and RX tasks $\mathcal{TX}$ and $\mathcal{RX}$, TT VCPUs $\mathcal{P}$, and Qbv streams $\mathcal{S}$. It derives a subset of dependent tasks $\mathcal{T}':=\{\tau|\forall \tau,\tau'\in\mathcal{T},\forall s\in\mathcal{S}: \tau\xrightarrow{s}\tau' \lor \tau'\xrightarrow{s}\tau\}$ and VCPUs $\mathcal{P}':=\{\rho|\forall \tau\in\mathcal{T}':\tau\in\mathcal{T}(\rho)\}$ to pass to the SMT-based dependency-aware network scheduler together with the sets $\mathcal{TX}$ and $\mathcal{RX}$. Our implementation of the SMT-based dependency-aware scheduler identifies groups of dependent schedulable entities (TT tasks, VCPUs, streams, TX and RX tasks) as in~\cite{DBLP:conf/rtns/RuhC24}, performs a necessary polynomial schedulability test~\cite{DBLP:journals/jacm/LiuL73}, and, if successful, jointly adds their resulting Constraints~\ref{const:release_time_deadline_wcet}--\ref{const:virtual_cpu_assignment} (and the 802.1Qbv Constraints~\cite{DBLP:conf/rtns/RuhC24}) to the SMT context. If the Z3~SMT solver returns true, the scheduler freezes the resulting segment offsets, segment sizes, and frame offsets and proceeds to the next group. Otherwise, the scheduler backtracks, combining groups to check a larger SMT context.

If the SMT-based dependency-aware scheduler finds schedule tables for $\mathcal{T}'$, $\mathcal{S}$, $\mathcal{TX}$, $\mathcal{RX}$, and $\mathcal{P}'$, it uses the resulting busy intervals $\mathcal{I}^c:=\{[z_{ij}.\phi,z_{ij}.\phi+z_{ij}.L]|\forall \rho_i\in\mathcal{P}'(c),\forall z_{ij}\in\mathcal{Z}_i\}$ and the remaining free TT tasks $\mathcal{T}\setminus\mathcal{T}'$ and VCPUs $\mathcal{P}\setminus\mathcal{P}'$ as an input to an EDF simulation of core $c$. The simulation of a core $c$ maintains a ready list dispatching TT tasks' jobs according to an EDF policy. If the simulation dispatches a job $J_{mn}\in\mathcal{J}_m$ of a free TT task $\tau_{m}\in\mathcal{T}\setminus\mathcal{T}'$ on a VCPU $\rho_k$, it differentiates two cases. In the case $\rho_k$ is already running, the simulation accounts for the task context switch time $\delta_\tau(\rho_k)$. If $\rho_k$ is not running, the simulation additionally accounts for the VCPU context switch time $\delta_\rho(c)$. Once context switch times have been accounted for, the simulation tracks a jobs runtime until the next preemption. When the simulation preempts a task job and/or VCPU, it creates segments $x_{mno}$ and $z_{kl}$. Before each busy interval $[z_{ij}.\phi, z_{ij}.\phi+z_{ij}.L]\in\mathcal{I}^c$ at simulation time $z_{ij}.\phi - \delta_{\rho}(c)$, the simulation checks if the VCPU $\rho_k\in\mathcal{P}\setminus\mathcal{P}'$ currently active within the simulation matches the busy interval's VCPU $\rho_i\in\mathcal{P}'$. If the VCPUs differ, the simulation preempts the currently active VCPU $\rho_k$ accounting for the VCPU preemption overhead $\delta_{\rho}(c)$ and resumes the simulation after the busy interval at time $z_{ij}.\phi+z_{ij}.L$. Otherwise, the simulation continues until time $z_{ij}.\phi$ to then preempt the currently active task job for the duration $z_{ij}.L$ of the busy interval. If the EDF simulation detects a deadline miss, it adds the corresponding free task and its VCPU to a set of incomplete tasks $\mathcal{T}^*$ and VCPUs $\mathcal{P}^*$ and continues the simulation.

\begin{figure*}[t!]
\centering
	\includegraphics[width=0.9\textwidth,trim={0.35cm 0.6cm 0.35cm 1.05cm},clip]{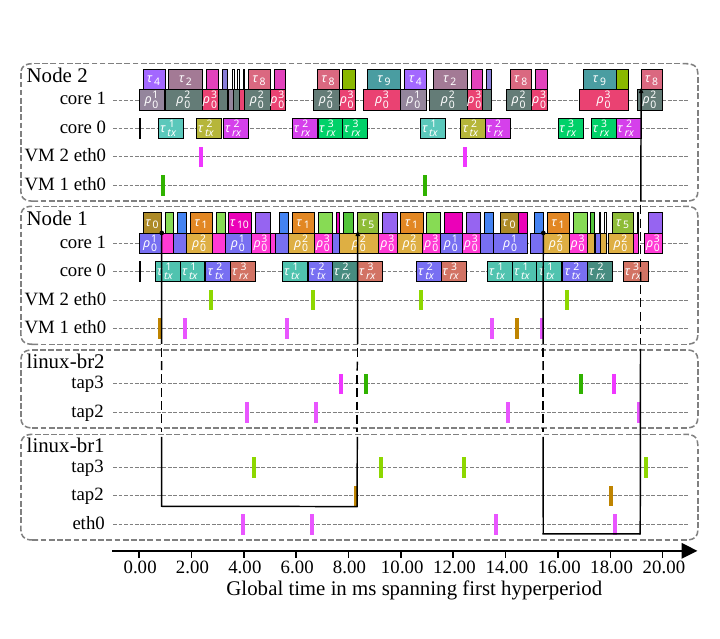}
	\caption{Example of a correct schedule}
	\label{fig:example_schedule}
\end{figure*}

When the EDF simulation on all cores returns and there are incomplete free TT tasks $\mathcal{T}^*$ and VCPUs $\mathcal{P}^*$, our scheduler adds them to the sets of dependent tasks $\mathcal{T}'=\mathcal{T}'\cup \mathcal{T}^*$ and VCPUs $\mathcal{P}'=\mathcal{P}'\cup \mathcal{P}^*$ and invokes the SMT-based dependency-aware scheduler again. The SMT-EDF scheduler alternates between the SMT-based dependency-aware scheduler and the EDF simulation until either there are no more incomplete free tasks, or, in the worst-case, it is $\mathcal{T}'=\mathcal{T}$ and $\mathcal{P}'=\mathcal{P}$ and the SMT-based dependency-aware scheduler solves the entire original problem. However, in the average case EDF-SMT scheduler exhibits greatly reduced runtime compared to the SMT-based dependency-aware scheduler.

Fig.~\ref{fig:example_schedule} shows a \emph{correct} schedule according to our Constraints~\ref{const:release_time_deadline_wcet}--\ref{const:rx_task_constraint} (and the network Constraints in~\cite{DBLP:conf/rtns/RuhC24}) synthesized by our SMT-EDF scheduler for a system with $2$ virtualized end systems, each hosting four VMs. One privileged VM per node runs the TX and RX tasks of a virtual TT network on a pinned single VCPU (core~$0$). These VMs access a passthrough NIC linking them. Three user VMs, each with one VCPU, share core~$1$. In total, the setup has $6$ user VMs hosting $15$ tasks, with $10$ communicating via $5$ Qbv streams. The data flow for one internal and one external stream is indicated by black arrows.

\subsection{A Time-Triggered Hypervisor}
\label{ssec:time_triggered_hypervisor}
We implemented a TT hypervisor with TT VCPU scheduling and TT VirtIO virtual networking based on the ACRN hypervisor~\cite{DBLP:conf/vee/LiXRD19} for x86-64 architectures. We chose ACRN over other open-source type-1 hypervisors, such as Xen~\cite{DBLP:conf/sosp/BarhamDFHHHN03}, or a type-2 hypervisor such as the Linux KVM, due to an existing clock synchronization infrastructure~\cite{Ruh21,DBLP:conf/dsn/RuhSF23} for the ACRN hypervisor that grants the hypervisor access to a global time base to align TT VCPU with task and network scheduling, a facility missing in Xen. In Linux KVM, a global time is available to the host through \lstinline|CLOCK_REALTIME| and the host and guests can use the same user space clock synchronization stack as in~\cite{Ruh21,DBLP:conf/dsn/RuhSF23}. Furthermore, several implementations of TT scheduling for the Linux kernel~\cite{Gala2021,Karachatzis23,gala2023} already exist. However, Linux KVM relies on QEMU~\cite{DBLP:conf/usenix/Bellard05} to configure VirtIO networking via \lstinline|vhost| in-kernel or run it directly in user space. 
Note that we did not consider purely static partitioning hypervisors, such as Jailhouse~\cite{DBLP:journals/corr/RamsauerKLM17} or Bao~\cite{DBLP:conf/hipeac/MartinsTSBP20}, since they do not support VCPU scheduling by design.

\subsubsection{TT VCPU Scheduling}
Officially, the ACRN hypervisor supports fixed priority (FP) scheduling, a borrowed virtual time~(BVT)~\cite{10.1145/319344.319169} scheduler, and static partitioning of cores via a no-operation~(NOOP) scheduler similar to Xen's NULL scheduler. 
Configuring schedule tables for TT VCPUs at compile time is relevant when considering safety and security. However, for this proof-of-concept~(PoC), we preferred setting VCPU schedule tables during runtime via a hypercall, easing development and testing. During startup until the privileged VM sets a schedule table for TT VCPUs and enables TT operation, our implementation utilizes a configurable secondary scheduler (BVT or FP) to dispatch VCPUs. Once the privileged VM switches the hypervisor to TT operation via a hypercall, the hypervisor dispatches TT VCPUs according to the intervals configured in the per-core schedule tables.

Let $n$ denote the number of entries in a core's schedule table. A hypervisor schedule table entry at an index $0\leq i\le n$ contains its offset $\phi_i$, its size (or relative deadline) $L_i$, a pointer to its associated VCPU $\rho_i$ (or \lstinline|NULL| for idle slots), and a pointer to the next schedule table entry $i+1$. We utilize the x86-64 hardware platform's timestamp counter~(TSC) deadline timer to prevent VCPUs from overrunning and to execute the schedule table. For a timer invocation $j$ during hyperperiod $hp$, the hypervisor checks if the schedule table entry is a VCPU entry, an idle entry, or an alignment entry and proceeds accordingly:
\begin{description}[style=unboxed,leftmargin=0cm,font=\textsmallbullet\space]
    \item[VCPU Entry] The TT scheduler sets the current schedule table entry to $i=(j+1)\mod n$, updates the active VCPU pointer, and arms the TSC deadline timer to preempt VCPU execution next at TSC $tsc_{j+1}=tsc_j+L_{i}$.
    \item[Idle Entry] The TT scheduler invokes the secondary scheduler to dispatch regular VCPUs until the end of the idle slot at $tsc_{j+1}=tsc_j+L_{i}$.
    \item[Alignment Entry] Let the function $g:\mathbb{N}\rightarrow\mathbb{N}$ return the global time at local time $tsc_j\in\mathbb{N}$. The TT scheduler reads the global time $g(tsc_j)$ provided by the dependent clock~\cite{Ruh21,DBLP:conf/dsn/RuhSF23} to calculate the runtime deviation $\Delta=g(tsc_j) - (hp \cdot T_{HP} + \phi_i)$ from the schedule table. If it is $\Delta < 0$, it means the TT execution is early compared to the intended execution. Thus, the hypervisor must realign by delaying the next TSC deadline timer for $\Delta$ cycles $tsc_{j+1}=tsc_j+L_{i}+\Delta$. If instead it is $\Delta > 0$, the TT execution is late, so the hypervisor must shorten the next TSC deadline timer by $\Delta$ cycles $tsc_{j+1}=tsc_j+L_{i}-\Delta$.
\end{description}
Lastly, when $(j+1)\mod n=0$, the TT scheduler completes a hyperperiod incrementing $hp$ and wrapping back to entry $0$. Note that if the misalignment exceeds the table entry size $|\Delta|>L_{i}$, the hypervisor must handle the resulting exception, and the TT scheduler must enter a predefined safe state. We can model alignment intervals on the hypervisor and guest OS levels as special TT tasks or VCPUs. Here, we pessimistically dimension alignment intervals and let the SMT-EDF scheduler place them periodically per core and VM. For example, in Fig.~\ref{fig:example_schedule}, we can identify alignment intervals on the hypervisor-level of node $2$ at time $0.0\,ms$ and at the guest-OS level at times $3.6\,ms$, $3.8\,ms$ and $4.0\,ms$.

\subsubsection{IEEE\hspace{0.1cm}802.1Qbv VirtIO Network}
We extended the ACRN DM's VirtIO network implementation to support up to eight transmit-receive \lstinline|virtqueue| pairs through multi-queue TAP devices. A \lstinline|virtqueue| is a shared memory region with a ring buffer between the privileged and a user VM. We map each TX and RX \lstinline|virtqueue| to a distinct \lstinline|pthread| that implements the TX and RX threads described in Sec.~\ref{ssec:virtual_network_model}. We acknowledge that the Linux kernel's \lstinline|vhost| virtual networking backend, also available in ACRN's privileged VM, already supports multi-queue operation serving TX and RX \lstinline|virtqueues| by distinct kernel threads. However, we found an implementation via \lstinline|pthreads| in the ACRN DM more straightforward for a PoC. The privileged VM in our setup runs a time-triggered Linux kernel~\cite{Karachatzis23}, enabling it to dispatch TX and RX \lstinline|pthreads| of IEEE\,802.1Qbv egress and ingress queues according to schedule tables generated by our EDF-SMT scheduler as described in Sec.~\ref{ssec:time_triggered_hypervisor}, enforcing the virtual network's IEEE\,802.1Qbv quasi-compliant operation. We say quasi-compliant, since to a user VM, the virtual multi-queue NIC is transparent and, presuming a GNU/Linux guest, the OS can configure a TAPRIO qdisc as it would for a passthrough physical NIC. Nevertheless, the TT privileged VM only provides timing guarantees on the corresponding virtual link if, in addition, a correctly aligned (c.f.~Sec.~\ref{ssec:virtual_network_constraints}) schedule table entry for the \lstinline|virtqueue|'s TX task exists. Furthermore, contrary to the IEEE\,802.1Qbv standard that enforces strict timing on the egress queue only, for strict timing guarantees in our TT virtual network, we also require a schedule table entry for the \lstinline|virtqueue|'s RX task; however, its configuration and existence remain transparent to a receiving VM.

The TT operation of TX and RX \lstinline|virtqueues| removes notification overheads compared to the regular, non-TT operation. In regular operation, a sending user VM's VirtIO frontend driver notifies the target virtqueue of the new data, resulting in a VM exit and the hypervisor injecting a virtual interrupt into the privileged VM. In the privileged VM, the interrupt invokes the DM to signal the corresponding TX \lstinline|pthread| to read and forward data from the \lstinline|virtqueue|. Similarly, on the receive path in regular operation, the Linux bridge queues frames in the target VM's TAP device, resulting in a callback that notifies the corresponding \lstinline|virtqueue|'s RX \lstinline|pthread|. The RX \lstinline|pthread| writes frames from the TAP device to its RX \lstinline|virtqueue| and notifies the target VM via a virtual interrupt once completed. In contrast, in TT operation, we disable \lstinline|virtqueue| notifications on the send and receive paths since we know a priori by the schedule tables when user VMs send and receive frames, rendering notifications moot.

\section{Experimental Evaluation}
\label{sec:evaluation}
Our experimental evaluation of the proposed time-triggered virtual networking is two-part. First, we evaluate the schedulability and runtime of the SMT-EDF offline scheduler for an increasing number of VMs with time-triggered VCPUs, considering inter-VM communication and external communication between VMs of two virtualized end systems. By increasing the number of VMs connected to the software bridge in a privileged VM, we automatically also increase the number of virtual links. Secondly, we present our testbed consisting of an industrial edge computing device running TT ACRN and time-triggered GNU/Linux in the privileged and user VMs executing generated task, stream, virtual network, and VCPU schedule tables. We report our findings, executing TT tasks with non-trivial dependencies in GNU/Linux VMs on the virtualized end system.

\subsection{Schedulability and Runtime}
\subsubsection{System Parameters}
\label{sssec:system_parameters}

Before we could generate schedules, we needed to to instantiate the system parameters introduced as part of the model and the correctness constraints (c.f. Sec.~\ref{sec:system_model} and \ref{sec:correctness_constraints}), namely, the task context switch overhead $\delta_{\tau}(\rho)$, the VCPU context switch overhead $\delta_{\rho}(c)$, the TT TX and RX tasks' TX and RX delays $\delta^{tx}$ and $\delta^{rx}$, the guest OS network stack's TX and RX overheads $o^{tx}$ and $o^{rx}$, and the clock synchronization precision $\Pi$. The target hardware platform is a commercial edge computing device featuring an Intel Atom E3950 multicore processor with four cores at $1.6\,GHz$ each and $8\,GiB$ main memory.

We measured the task context switch time $\delta_{\tau}(\rho)$ forcing two high priority processes to repeatedly preempt each other by alternately reading from blocking pipes~\cite{DBLP:conf/usenix/Ousterhout90a}. In between preemptions, each process performed strided memory reads/and writes to incur worst-case cache interference~\cite{DBLP:conf/expcs/LiDS07}. The measurement tasks took timestamps before each pipe read to calculate the task context switch time. We could not apply the same methodology to measure the VCPU context switch time $\delta_{\rho}(c)$ since there exists no standard mechanism to force alternating preemptions between the VCPUs of a hypervisor. Therefore, we instrumented a x86-64 model specific register (MSR) to cause a VM exit upon read access. A simple dispatch logic in the MSR access VM exit handler forced a VCPU context switch. To measure the VCPU context switch time, two VMs with one VCPU each shared a processor core. Each VCPU executed a simple measurement task, that, again, performed strided memory reads/and writes to incur worst-case cache interference and then read the instrumented MSR to force a VCPU preemption and context switch to the partner measurement VM's VCPU. The measurement task took timestamps before each MSR read to calculate the VCPU context switch time. For task and VCPU context switch times we collected $2\cdot 10^5$ data points each. To determine $\delta^{tx}$, $\delta^{rx}$, $o^{tx}$, and $o^{rx}$, we generated tables for two TT tasks executing on distinct VCPUs sharing cores and communicating via a Qbv stream. This presented a mutual dependency, since schedule generation required us to already know said parameters. Therefore, we presumed overly pessimistic values of $\delta^{tx*}=1\,ms$, $\delta^{rx*}=1\,ms$, $o^{tx*}=1\,ms$, and $o^{rx*}=1\,ms$. We further instantiated the task and VCPU context switch times $\delta_{\tau}(c)=115\,\mu s$ and $\delta_{\rho}(c)=165\,\mu s$ to the max. measured value by the method described before and the precision to $\Pi=0$ since for this measurement VMs shared a host. We executed the resulting set of schedule tables for $30$\,minutes and measured the actual values of $\delta^{tx}$, $\delta^{rx}$, $o^{tx}$ and $o^{rx}$ instrumenting the ACRN DM. We synchronized the dependent clock~\cite{Ruh21} of the privileged VMs on the two industrial edge computing devices using LinuxPTP. We derived the upper bound on the precision $\Pi=43\,ns$ as in~\cite{Ruh21}. Furthermore, we report the measured synchronization precision $\Pi^*$ with $\Pi^*\leq\Pi$. Fig.~\ref{fig:system_parameters} summarizes the measurement results. The box plots' whiskers capture the $Q\,0.05$ and $Q\,0.95$ quantiles, the red line the median, and the red $\times$ marks the max measured value.

Our PoC based on GNU/Linux VMs and an uncertified open-source hypervisor has only limited applicability for hard real-time systems~\cite{Reghenzani19}. However, using GNU/Linux may be viable in safety-critical systems considering certain tradeoffs~\cite{1137388,deoliveira_et_al:LIPIcs.ECRTS.2020.9}. For instance, one such trade-off arises from our choice of the system parameters to generate schedules. Using the median or $Q,0.95$ measured values leads to higher system utilization but may result in occasional deadline misses, whereas using the maximum measured value (or an analytical worst-case) ensures no deadline misses at the expense of schedulability. Note that in practice, there are varying system requirements concerning the rate of deadline misses we deem tolerable. In industrial systems~\cite{ademaj2019industrial,IIC_TSN_Traffic_2018}, periodic critical events map to either isochronous TSN streams, which cannot miss any deadlines, or to cyclic TSN streams, which can tolerate $1-4$ consecutive frame drops. Typically, we design industrial and automotive systems to tolerate some degree of deadline misses.

\subsubsection{Experimental Results}
\label{sssec:experimental_setup}
\begin{wrapfigure}[16]{h}{0.5\textwidth}
\vspace{-1.5cm}
\includegraphics[width=0.49\textwidth,trim={0.3cm 0.3cm 0.4cm -0.61cm},clip]{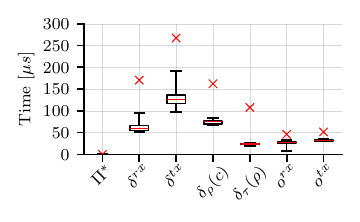}
    	\caption{System overheads}
    	\label{fig:system_parameters}
\end{wrapfigure}

Our schedulability experiments for synthetic task sets and streams use parameters from two real-world automotive benchmarks, referred to as TTTech~\cite{FinziRTAS24}~(c.f.~Sec.\,IV.B) and WATERS~\cite{Kramer15}. For each benchmark, we evaluate the schedulability and runtime when using the $Q\,0.95$ system parameters compared to the max measured system parameters (c.f.~Fig.~\ref{fig:system_parameters}) for an increasing number of VMs and streams on two virtualized end systems whose privileged VMs have access to passthrough NICs with a direct physical link.

For both benchmarks, we randomly generated $10$, $20$, and $30$ IEEE\,802.1Qbv streams selecting stream sizes from the set $\{1,\allowbreak 2,\allowbreak 4,\allowbreak 8,\allowbreak 16,\allowbreak 32,\allowbreak 64,\allowbreak 3000\}\,$byte with distribution $\{0.35, 0.49, 0.13, 0.008, 0.013, 0.005, 0.002, 0.002\}$ resulting in $20$, $40$, $60$, and $80$ communicating TT tasks. For the TTTech benchmark~\cite{FinziRTAS24}, we generated a total $120$ TT tasks with periods randomly selected from the set $\{5,10,20,40,80\}\,ms$ with a distribution of $\{0.09166, 0.2666,0.125,0.19166, 0.325\}$, i.e., if there are $n\in\{20,40,60\}$ communicating tasks we additionally generate $(120-n)$ free TT tasks. We randomly picked the WCETs utilizing the \lstinline|Randfixedsum| algorithm~\cite{emberson_taskgen,emberson2010techniques} to approximate the per-core utilization of approximately $50\%$ including context switch overheads. For the WATERS benchmark~\cite{Kramer15}, we randomly selected TT tasks' periods from the set $T:=\{1,2,5,10,20,50,100,200,1000\}\,ms$ with distribution $\{0.03, 0.02,0.02, 0.25,0.25, 0.03,0.2, 0.01, 0.04\}$. Given a task's random period $T_i\in T$ at index~$i$, we select its corresponding average-case execution time at index~$i$ from the set $ACETs=\{5, 4.2,11.04, 10.09,8.74, 17.56,10.53, 2.56,0.43\}\mu s$ and its minimum and maximum WCET factors $f_{min}$ and $f_{max}$ from the sets $F_{min}=\{1.3, 1.54,\allowbreak 1.13, 1.06,\allowbreak 1.06, 1.13,\allowbreak 1.02, 1.03, 1.84\}$ and $F_{max}=\{29.11, 19.04,18.44,30.03, 15.61,7.76, 8.88,4.9, 4.75\}$, respectively. Then, we pick a random WCET factor $f\in[f_{min},f_{max}]$ and calculate the task's WCET $C_i=f\cdot ACETs[i]$. We assigned each generated task to a VCPU of a core with spare capacity, and repeated the procedure until it reaches the per-core target utilization of approximately $50\%$ including context switch overheads. For a given pair $\tau_i,\tau_j$ TT tasks communicating via a stream $s_k$, we assume that they share a period, i.e., we do not consider multi-rate dependencies, and set their end-to-end requirement equals their period $E(\tau_i\xrightarrow{s_k}\tau_j)=T_i=T_j$.

\begin{figure}[t!]
\centering
	\includegraphics[width=0.85\textwidth,trim={0.45cm 0.8cm 0.5cm 0.3cm},clip]{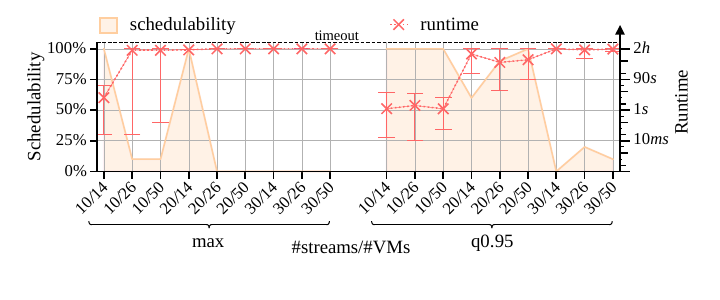}
	\caption{Schedulability and runtime using max. and $q0.95$ system parameters}
	\label{fig:exp_sched}
\end{figure}

For each schedulability experiment's configuration, we generated and scheduled $10$ random task sets. We assumed a realistic configuration of the GNU/Linux privileged and user VMs. Therefore, for each target device, we statically allocated core~$0$ to host its privileged VM's VCPU with the TT TX and RX tasks. We configured each user VM $i$ with $2$ VCPUs so that VCPU $\rho^i_0$ acted as a bookkeeping core isolating the TT VCPU $\rho^i_1$\footnote{kernel cmd. line parameters: \lstinline|rcu_nocb=1 rcu_nocb_poll nohz_full=1 isolcpus=1|}. Bookkeeping VCPUs $\{\rho^i_0\}$ shared core $1$ scheduled with BVT whereas TT VCPUs $\{\rho^i_1\}$ were equally distributed to the remaining cores $2$ and $3$. To counter a high schedule generation runtime, $14$ EDF-SMT scheduler instances solved experiment configurations in parallel on an AMD\,Ryzen\,7\,57003D with $8$~physical cores ($16$ logical cores) at $2.64\,GHz$ each running Kubuntu~$25.04$ with Linux kernel~v6.14.0 accessing $128\,GiB$ of RAM. We set an EDF-SMT scheduler timeout of $2\,h$.

Fig.~\ref{fig:exp_sched} illustrates the
schedulability and runtime results for the TTTech benchmark~\cite{FinziRTAS24}. The left y-axis we shows the schedulability, i.e., how many instances were schedulable until the time-out, and the right, logarithmic y-axis shows the synthesis runtime. The x-axis displays the system size in terms of number of streams and number of VMs. We can see that when comparing max. and mean used delays during schedule synthesis, for max. parameters the schedulability starts declining already for smaller system sizes due to the high delay values, whereas with the q95 parameters, the schedulability remains high for larger system sizes. Both the number of streams and number of VMs influence the schedulability and runtime. The WATERS benchmark timed out for all instances mainly due to the big difference between the lowest and the highest periods and is not shown here. For such systems, either a separation on different physical cores based on period values~\cite{7000667}, period transformation~\cite{7176034}, or a pure heuristic-based solution are preferable (e.g.~\cite{DBLP:journals/rts/SyedF19}). 

\begin{wrapfigure}[20]{h}{0.5\textwidth}
\vspace{-0.5cm}
\includegraphics[width=0.49\textwidth,trim={0.3cm 0.3cm 0.3cm 0.3cm},clip]{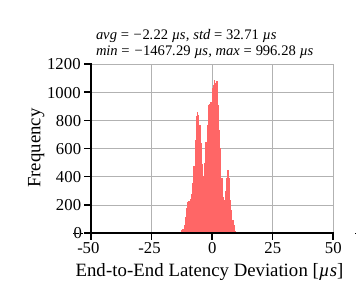}
    	\caption{End-to-end latency deviation from schedule tables for $5$ streams.}
    	\label{fig:end2end-1}
\end{wrapfigure}
Our TT testbed consists of single edge computing device (c.f.~Sec.~\ref{sssec:system_parameters}) running ACRN $v2.4$ with TT VCPU scheduling, the privileged VM hosting the TT virtual networks' TX and RX tasks, and TT user VMs. The number of user VMs hosted on the edge computing device is limited by the available persistent memory to store VM images. Currently, we are hosting $6$ user VMs (as described in Sec.~\ref{sssec:experimental_setup}) with $1$ bookkeeping VCPU and $1$ TT VCPU each. We assigned the bookkeeping VCPUs to core $1$ and the TT VCPUs to cores $2$ or $3$, respectively, resulting in a total of $6$ bookkeeping VCPUs on core $1$ and $3$ TT VCPUs each per cores $2$ and $3$. Exploring the schedule tables, we found that the GNU/Linux-based implementation requires additional system-specific constraints to correctly handle overheads resulting from the Linux kernel's software-interrupt handling and the read-copy-update (RCU) subsystem. For now these unpredictable overheads in the GNU/Linux prototype resulted in an increasing number of failed streams. For up to $5$ TT streams, $60$ free TT tasks resulting in a total target utilization of $50\%$ per TT core, and interfering \lstinline|iperf| network traffic between VMs, we found $95\%$ of stream jobs successfully completing. Fig.~\ref{fig:end2end-1} shows the deviation of the measured end-to-end latency from the end-to-end latency dictated by the schedule tables. Despite outliers expected in a GNU/Linux-based implementation, we can see that the measured end-to-end latency of the $5$ TSN streams follows the expected end-to-end latency tightly only deviating by $\pm12\,\mu s$.

\section{Conclusion}
\label{sec:conclusion}

We have proposed a VirtIO-based virtual network that realizes IEEE 802.1Qbv quasi-compliant inter-VM communication for time-triggered hypervisors, providing predictable timing for critical messages. We presented our VirtIO-based virtual network architecture and derived network-level scheduling constraints that emerge from the proposed solution's properties. Schedulability and schedule synthesize runtime experiments for two benchmark task sets have highlighted the difficulty of generating global schedule tables considering realistic systems parameters. To the authors knowledge, our PoC utilizing the ACRN hypervisor is the first demonstrating tight end-to-end latency guarantees of TT tasks hosted on TT VCPUs communicating via a TT virtual network. Addressing present drawbacks of our PoC requires a close interplay of (operating) system engineering and modeling resulting system-specific constraints.

\section*{Acknowledgements}
The work is co-funded by the Chips Joint Undertaking (Chips JU) and National Authorities (FFG Grant Agreement No. 50307277) under grant agreement No. 101139789 (HAL4SDV project). 
\begin{figure}[h!]
    \centering
	\includegraphics[width=0.3\textwidth]{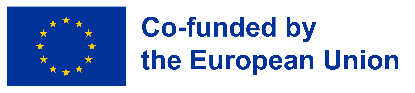}
    \includegraphics[width=0.3\textwidth]{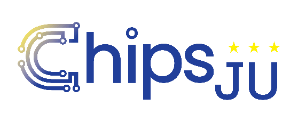}
\end{figure}

\bibliographystyle{unsrtnat}
\bibliography{references}

\clearpage


\section*{Supplementary material}

# Errata for "Towards Virtual Networking for a Time-Triggered Hypervisor" (RTNS 2025)

Jan Ruh Silviu S. Craciunas Jan Korecky

October 13, 2026

This document corrects errors in the formal system model and correctness constraints of the paper "Towards Virtual Networking for a Time-Triggered Hypervisor," published in the *Proceedings of the 33rd International Conference on Real-Time Networks and Systems (RTNS 2025)*. The errors concern the formal statement of the model and constraints. The schedule synthesis implementation used in the experimental evaluation enforced the intended semantics, so the experimental results and the conclusions of the paper are unaffected. Constraints 1–5 and 8–10 of the paper recapitulate constraints of the RTNS 2024 paper "End-to-End Schedulability of Virtualized Distributed Time-Triggered Systems" (DOI: 10.1145/3696355.3696364) and inherit some of the errors corrected in the errata of that paper.

## Erratum 1 — Typographical correction, TX/RX task definitions, Section 3.2

Two corrections apply to the definition of the virtual-network backend tasks.

**(a)** The RX task must be a member of the RX task set, $\tau^{rx}_{v,i} \in \mathcal{RX}^v(\rho_p)$, not of $\mathcal{TX}^v(\rho_p)$ as published (likewise in the quantifier prefix of Constraint 12, cf. Erratum 4).

**(b)** The period of an RX task must be derived from the received streams, $T^v_i = \mathrm{lcm}\{T_x \mid \forall s_x \in \mathcal{S}^{rx}_{v,i}\}$, not from $\mathcal{S}^{tx}_{v,i}$ as published.

## Erratum 2 — Joint unsatisfiability of Constraints 5, 6, and 7, Section 4.2

The first conjunct of Constraint 5 is an alternative to Constraint 6, while its second conjunct is an alternative to Constraint 7. These alternatives must not be applied jointly, but this clarification is missing in the paper. As published, all three constraints are stated over the same universally quantified domain and when applied jointly, they are unsatisfiable. In the implementation, Constraint 6 is only applied if a stream's first link is a virtual link, and Constraint 7 is only applied if a stream's last link is a virtual link. Accordingly, the first or second conjunct of Constraint 5 does not apply in the respective case. The paper therefore needs to include the following clarification after Constraint 7:

> *Constraints 5–7 apply mutually exclusively, depending on the type of a stream's first and last link. If the first link $(a, x)$ of a stream $s_h$ is a virtual link, Constraint 6 applies and the first conjunct of Constraint 5 is omitted. Otherwise, the first conjunct of Constraint 5 applies and Constraint 6 is omitted. Analogously, if the last link $(y, b)$ of $s_h$ is a virtual link, Constraint 7 applies and the second conjunct of Constraint 5 is omitted. Otherwise, the second conjunct of Constraint 5 applies and Constraint 7 is omitted. Hence, for every stream, exactly one of Constraint 5's first conjunct and Constraint 6 holds, and exactly one of Constraint 5's second conjunct and Constraint 7 holds.*

## Erratum 3 — Helper set $\mathcal{R}^v_{ijk}$, Section 4.4

Three corrections apply to the helper set for the TX/RX task job segment size constraints.

**(a)** The job-instance mapping is computed as $j = \lfloor (g \cdot T_f)/T^v_i \rfloor$ using the TX/RX task period $T^v_i$, but the two containment inequalities add the instance offset as $j \cdot T_i$, where $i$ is a queue index and $T_i$ is undefined. Both occurrences must read $j \cdot T^v_i$.

**(b)** The published segment-internal reservation

$$\delta(x^v_{ijk}) = \begin{cases} \delta^{tx}, & \text{if } x^v_{ijk} \text{ is a TX task job segment} \\ -\delta^{rx}, & \text{if } x^v_{ijk} \text{ is a RX task job segment} \end{cases}$$

carries a sign error: with $-\delta^{rx}$, the upper containment bound $f_{fgh}.\phi + f_{fgh}.L + g \cdot T_f \leq x_{ijk}.\phi + x_{ijk}.L - \delta(x^v_{ijk}) + j \cdot T^v_i$ permits frames of an RX segment to complete $\delta^{rx}$ *after* the segment ends, contradicting Constraint 12, which reserves $\delta^{rx}$ per stream job *inside* the segment. The definition must read $\delta(x^v_{ijk}) = \delta^{rx}$ for RX task job segments.

**(c)** The helper set is defined only for TT TX tasks. The definition for TT RX tasks is analogous. We can write:

$$\begin{aligned} \mathcal{R}^{v,x}_{ijk} := \{ J^x_{fg} | \forall s_f \in \mathcal{S}^x_{v,i}, \forall J^x_{fg} \in \mathcal{J}^x_{v,i}, \forall f_{fgh} \in \mathcal{F}^x_{fg} : \\ j = \left\lfloor \frac{g \cdot T_f}{T^v_i} \right\rfloor \wedge (f_{fgh}.\phi + g \cdot T_f \geq x_{ijk}.\phi + j \cdot T^v_i) \\ \wedge (f_{fgh}.\phi + f_{fgh}.L + g \cdot T_f \leq x_{ijk}.\phi + x_{ijk}.L - \delta(x^v_{ijk}) + j \cdot T^v_i) \} \end{aligned}$$

with $x \in \{tx, rx\}$.

## Erratum 4 — Constraints 11 and 12 (TX/RX Task Job Segment Size), Section 4.4

There are some typos and spurious variables that can be removed, hence the corrected constraints read:

**Constraint 11** (TX Task Job Segment Size Constraint)**.**

$$\begin{aligned} \forall n \in \mathcal{V}, \forall v \in \mathcal{V}^n, \forall c \in \mathcal{C}^n, \forall \rho_p \in \mathcal{P}(c), \forall \tau^{tx}_{v,i} \in \mathcal{TX}^v(\rho_p), \forall J^{tx}_{ij} \in \mathcal{J}^{tx}_i, \forall x^v_{ijk} \in \mathcal{X}^{tx}_{ij} : \\ x^v_{ijk}.L \geq \delta_\tau(\rho_p) + \sum_{\forall J^{tx}_{fg} \in \mathcal{R}^{v,tx}_{ijk}} \Big( \delta^{tx} + \sum_{\forall f_{fgh} \in \mathcal{F}^{tx}_{fg}} f_{fgh}.L \Big) \end{aligned}$$

**Constraint 12** (RX Task Job Segment Size Constraint)**.**

$$\begin{aligned} \forall n \in \mathcal{V}, \forall v \in \mathcal{V}^n, \forall c \in \mathcal{C}^n, \forall \rho_p \in \mathcal{P}(c), \forall \tau^{rx}_{v,i} \in \mathcal{RX}^v(\rho_p), \forall J^{rx}_{ij} \in \mathcal{J}^{rx}_i, \forall x^v_{ijk} \in \mathcal{X}^{rx}_{ij} : \\ x^v_{ijk}.L \geq \delta_\tau(\rho_p) + \sum_{\forall J^{rx}_{fg} \in \mathcal{R}^{v,rx}_{ijk}} \Big( \delta^{rx} + \sum_{\forall f_{fgh} \in \mathcal{F}^{rx}_{fg}} f_{fgh}.L \Big) \end{aligned}$$

## Erratum 5 — Typographical correction, Section 4.2

In the paragraph preceding Constraint 5, "Furthermore, he transmission schedule of the stream $s_k$ aligns along its route" must read "Furthermore, the transmission schedule of the stream $s_h$ aligns along its route" (the constraint binds $s_h$).



\end{document}